\documentclass[aps,prb,amsmath,amssymb,reprint,superscriptaddress,]{revtex4-2}

\usepackage{graphicx}
\begin{document}
\title{Boiling and cavitation caused by transient heat transfer in superfluid helium-4}

\author{Hamid Sanavandi}
\affiliation{National High Magnetic Field Laboratory, 1800 East Paul Dirac Drive, Tallahassee, Florida 32310, USA}
\affiliation{Mechanical Engineering Department, FAMU-FSU College of Engineering, Florida State University, Tallahassee, Florida 32310, USA}

\author{Mikai Hulse}
\affiliation{National High Magnetic Field Laboratory, 1800 East Paul Dirac Drive, Tallahassee, Florida 32310, USA}
\affiliation{Department of Physics, Florida State University, Tallahassee, Florida 32306, USA}

\author{Shiran Bao}
\affiliation{National High Magnetic Field Laboratory, 1800 East Paul Dirac Drive, Tallahassee, Florida 32310, USA}
\affiliation{Mechanical Engineering Department, FAMU-FSU College of Engineering, Florida State University, Tallahassee, Florida 32310, USA}

\author{Yuan Tang}
\affiliation{National High Magnetic Field Laboratory, 1800 East Paul Dirac Drive, Tallahassee, Florida 32310, USA}
\affiliation{Mechanical Engineering Department, FAMU-FSU College of Engineering, Florida State University, Tallahassee, Florida 32310, USA}

\author{Wei Guo}
\email[Corresponding: ]{wguo@magnet.fsu.edu}
\affiliation{National High Magnetic Field Laboratory, 1800 East Paul Dirac Drive, Tallahassee, Florida 32310, USA}
\affiliation{Mechanical Engineering Department, FAMU-FSU College of Engineering, Florida State University, Tallahassee, Florida 32310, USA}

\begin{abstract}
Superfluid helium-4 (He II) has been widely utilized as a coolant in various scientific and engineering applications due to its superior heat transfer capability. An important parameter required in the design of many He II based cooling systems is the peak heat flux $q^*$, which refers to the threshold heat flux above which boiling spontaneously occurs in He II. Past experimental and numerical studies showed that $q^*$ increases when the heating time $t_h$ is reduced, which leads to an intuitive expectation that very high $q^*$ may be achievable at sufficiently small $t_h$. Knowledge on how $q^*$ actually behaves at small $t_h$ is important for applications such as laser ablation in He II. Here we present a numerical study on the evolution of the thermodynamic state of the He II in front of a planar heater by solving the He II two-fluid equations of motion. For an applied heat flux, we determine the heating time beyond which the He II near the heater transits to the vapor phase. As such, a curve correlating $q^*$ and $t_h$ can be obtained, which nicely reproduces some relevant experimental data. Surprisingly, we find that there exists a critical peak heat flux $q^*_c$, above which boiling occurs nearly instantaneously regardless of $t_h$. We reveal that the boiling in this regime is essentially cavitation caused by the combined effects of the first-sound and the second-sound waves in He II. Based on this physical picture, an analytical model for $q^*_c$ is developed, which reproduces the simulated $q^*_c$ values at various He II bath temperatures and hydrostatic head pressures. This work represents a major progress in our understanding of transient heat transfer in He II.
\end{abstract}

\maketitle

\section{Introduction}
When saturated liquid $^4$He is cooled to below about 2.17 K, it undergoes a phase transition to the superfluid phase (known as He II)~\cite{Tilley-book}. Phenomenologically, He II can be regarded as a mixture of two miscible fluid components: an inviscid and zero-entropy superfluid and a viscous normal fluid that consists of thermal quasiparticles (i.e., phonons and rotons)~\cite{landau-1987}. He II has many unique thermal and mechanical properties. For instance, it supports two sound-wave modes: an ordinary pressure-density wave (i.e., the first sound) where the two fluids oscillate in phase, and a temperature-entropy wave (i.e., the second sound) where the two fluids oscillate oppositely~\cite{landau-1987}. Furthermore, heat transfer in He II is via an extremely effective counterflow mode~\cite{landau-1987}: the normal fluid carries the heat and moves away from the heat source at a mean velocity $v_n$=$q/\rho sT$ , where $q$ is the heat flux, $T$ is the He II temperature, $\rho$ and $s$ are respectively the total density and the specific entropy of He II; while the superfluid moves in the opposite direction at a mean velocity $v_s$=$-v_n\rho_n/\rho_s$ to ensure zero net mass flow (here $\rho_s$ and $\rho_n$ denote the densities of the superfluid and the normal fluid, respectively). When the relative velocity of the two fluids exceeds a small critical value~\cite{Vinen-1957-PRS-I}, a chaotic tangle of quantized vortex lines are created spontaneously in the superfluid, each carrying a quantized circulation $\kappa\simeq10^{-3}$ cm$^2$/s around its angstrom-sized core~\cite{donnelly-1991}. A mutual friction force between the two fluids then emerges due to the scattering of the thermal quasiparticles off the quantized vortices~\cite{vinen-1957_Proc.R.Soc.Lond.A_II}, which can lead to novel flow characteristics in both fluids~\cite{marakov-2015_Phys.Rev.B, Gao-2017-PRB, Gao-2018-PRB, Bao-2018-PRB,Mastracci-2018-PRF}.

Due to the superior heat transfer capability, He II has been widely utilized in various scientific and engineering applications for cooling devices such as superconducting magnets, superconducting accelerator cavities, and satellites~\cite{vansciver-2012}. Many of these applications involve high-flux transient heat transfer from a solid surface to He II, which is a complex process that involves the interplay of counterflow, second-sound emission, and vortex generation~\cite{SBaoTHTHeII}. When the heat flux is higher than a threshold value denoted as the peak heat flux $q^*$, boiling of He II can spontaneously occur on the heating surface. This peak heat flux $q^*$ is an important parameter needed in the design of many He II based cooling systems, which has been the subject of extensive experimental and numerical studies~\cite{kobayashi1980maximum,nemirovskii1989transient,tsoi1985boiling,danilchenko1989limit,wang1995criterion,shimazaki1998measurement,kondaurova2017influence,kondaurova2018dynamics}. A power-law dependance $q^*\propto t_h^{-n}$ of $q^*$ on the heating time $t_h$ was reported in literature, where the power index $n$ varies in the range of 0.25 to 0.5 depending on the magnitude of the applied heat flux and other experimental conditions such as the He II bath temperature $T_b$ and the hydrostatic head pressure $P_h=\rho g H$ (where $H$ is the He II depth and $g$ is the gravitational acceleration)~\cite{tsoi1985boiling,danilchenko1989limit,wang1995criterion,shimazaki1998measurement,kondaurova2017influence}. These results may give an intuitive expectation that very high $q^*$ may be achievable at sufficiently small $t_h$. However, knowledge on how $q^*$ actually behaves in the high heat flux and short heating time regime is limited. Such knowledge could benefit various research fields such as nanomaterial fabrication via laser ablation in He II~\cite{Latimer-2014-NL,Gordon-2018-LPL}.

In this paper, we present a numerical study of the $q^*$-$t_h$ correlation with an emphasis on the high heat-flux regime. By solving the He II two-fluid equations of motion coupled with the Vinen's equation for the vortex-line density~\cite{vinen-1957_Proc.R.Soc.Lond.A_II}, we examine the evolution of the thermodynamic state of the He II in front of a planar heater. For an applied heat flux $q$, we determine the heating time $t_h$ beyond which the He II near the heater surface transits to the vapor phase. The obtained $q^*$-$t_h$ curves show good agreement with relevant experimental data. Surprisingly, we find that there exists a critical peak heat flux $q^*_c$, above which boiling occurs nearly instantaneously regardless of $t_h$. Our analysis shows that the boiling in this regime is indeed heat-induced cavitation on the heater surface caused by combined effects of the first-sound and the second-sound waves in He II accompanying the heat transfer. We discuss an analytical model for evaluating $q^*_c$ and show that this model can nicely reproduce the simulated $q^*_c$ values at various He II bath temperatures and hydrostatic head pressures.

The paper is organized as follows. In Sec.~\ref{sec:model}, we outline the numerical model adopted in our study and briefly discuss some characteristic features of transient heat transfer in He II. In Sec.~\ref{sec:peak}, we explain how the $q^*$-$t_h$ correlation is determined at a given $T_b$ and the He II depth $H$. A comparison of the obtained $q^*$-$t_h$ curves with available experimental data is made, and the appearance of the unexpected critical peak heat flux $q^*_c$ is discussed. In Sec.~\ref{sec:variation}, we present a systematic study on how $q^*_c$ depends on $T_b$ and $H$. The underlying physical mechanism of $q^*_c$ is explained in Sec.~\ref{sec:explanation}. Finally, a brief summary is given in Sec.~\ref{sec:sum}.

\section{\label{sec:model}Numerical model}
There have been a variety of methods developed for studying heat transfer and flow dynamics in He II under the influence of quantized vortices. For a comprehensive review of these methods, one may refer to Refs.~\cite{Nemirovskii-1995-RMP,nemirovskii2020closure}. In this study, we adopt the two-fluid hydrodynamic model presented in our previous work~\cite{SBaoTHTHeII} to examine 1D heat transfer in He II from a planar heater. This model consists of the following conservation equations for the He II mass, momentum, and energy:
	\begin{gather}
		\frac{\partial\rho}{\partial t}+\mathbf{\nabla}\cdot(\rho \mathbf{v})=0,\label{eq:mass}\\
        \frac{\partial(\rho\mathbf{v})}{\partial t}+\mathbf{\nabla}(\rho_{s}v_{s}^2+\rho_{n}v_{n}^2)+\mathbf{\nabla}P=0,\label{eq:v}\\
		\frac{\partial{ \mathbf{v_s}}}{\partial t}+\mathbf{v_s}\cdot{\mathbf{\nabla}\mathbf{v_s}}+\mathbf{\nabla}\mu=\frac{\mathbf{F_{ns}}}{\rho_s},\label{eq:vs}\\
        \frac{\partial(\rho s)}{\partial t}+\mathbf{\nabla}\cdot(\rho s \mathbf{v_n})=\frac{\mathbf{F_{ns}}\cdot\mathbf{v_{ns}}}{T},\label{eq:energy}
    \end{gather}
where $\rho {\bf v}=\rho_{s}{\bf v_{s}}+\rho_{n}{\bf v_{n}}$ denotes the total momentum density, $P$ is the He II pressure, and $\mu$ is the chemical potential of He II. Considering the large relative velocity $v_{ns}=|\mathbf{v_{ns}}|=|\mathbf{v_{n}}-\mathbf{v_{s}}|$ of the two fluids at high heat fluxes, we also include the corrections to the thermodynamic properties of He II as proposed by Landau~\cite{landau-1987,Khalatnikov-book}:
\begin{gather}
	\mu(P,T,v_{ns})=\mu^{(s)}(P,T)-\frac{1}{2}\frac{\rho_n}{\rho}v_{ns}^2,\label{eq:mu}\\
    s(P,T,v_{ns})=s^{(s)}(P,T)+\frac{1}{2}v_{ns}^2\partial(\rho_n/\rho)/\partial{T},\label{eq:s}\\
    \rho(P,T,v_{ns})=\rho^{(s)}(P,T)+\frac{1}{2}\rho^2v_{ns}^2\partial(\rho_n/\rho)/\partial{P};\label{eq:rho}
\end{gather}
where all the static properties (i.e., with superscript $^{(s)}$) are extracted from the Hepak dynamic library~\cite{hepak-2005}. Typically, the corrections amount to no more than a few percent of the property values in static He II.

The Gorter-Mellink mutual friction $\mathbf{F_{ns}}$ per unit fluid volume depends on the vortex-line density $L$ (i.e., length of vortices per unit volume) and $\mathbf{v_{ns}}$ as~\cite{vinen-1957_Proc.R.Soc.Lond.A_III}:
\begin{equation}
	\mathbf{F_{ns}}=\frac{\kappa}{3}\frac{\rho_s\rho_n}{\rho}B_LL\mathbf{v_{ns}}
	\label{eq:Fns}
\end{equation}
where $B_L$ is a temperature-dependent mutual friction coefficient~\cite{Donnelly-1998-JPCRD}. To determine $\mathbf{F_{ns}}$, we need to know the spatial and temporal evolution of $L(\mathbf{r},t)$. For this purpose, we adopt the Vinen's equation~\cite{vinen-1957_Proc.R.Soc.Lond.A_II}:
\begin{equation}
\frac{\partial L}{\partial t}+{\bf\nabla}\cdot({\bf v_L}L)=\alpha_V|{\bf v_{ns}}|L^{\frac{3}{2}}-\beta_VL^2+\gamma_V|{\bf v_{ns}}|^{\frac{5}{2}}\label{eq:L}
\end{equation}
where $\alpha_V$, $\beta_V$, and $\gamma_V$ are temperature-dependent phenomenological coefficients~\cite{vinen-1957_Proc.R.Soc.Lond.A_II}. We use the values recommended by Kondaurova \emph{et al.} for these coefficients since they give simulation results that agree well with experimental observations~\cite{kondaurova-2017_J.LowTemp.Phys.}. The term $\mathbf{\nabla}\cdot(\mathbf{v_L}L)$ accounts for the drifting of the vortices~\cite{schwarz-1988_Phys.Rev.B, nemirovskii-2019_LowTemp.Phys.}, where we take the drift velocity $\mathbf{v_L}$ to be the local superfluid velocity $\mathbf{v_s}$~\cite{vinen-1957_Proc.R.Soc.Lond.A_II}. Nonetheless, we find that this drifting effect is negligible in highly transient heat transfer processes~\cite{SBaoTHTHeII}. The above model represents a coarse-grained description of the two-fluid hydrodynamics, since the action of individual vortices on the normal fluid is not resolved~\cite{Yui-2020-PRL, Mastracci-2019-PRF}. Nonetheless, when the vortex-line density is relatively high, this model was shown to describe non-isothermal flows in He II very well~\cite{Sergeev-2019-EPL, SBaoTHTHeII}.

\begin{figure*}[t]
		\centering
		\includegraphics[width=0.98\linewidth]{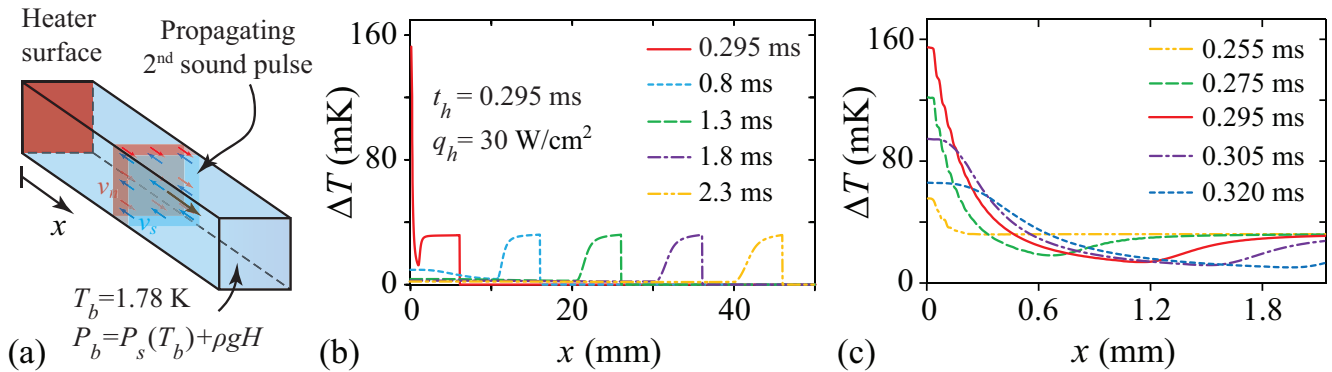}
		\caption{(a) A schematic diagram showing the transient heat transfer process from a planar heater located at $x=0$. (b) Simulated curves showing the temperature increment $\Delta T=T-T_b$ in He II at various times when a heat pulse of $q_h=30$ W/cm$^2$ with a duration $t_h=0.295$~ms is applied. The bath temperature is $T_b=1.78$ K, and a He II depth $H=1$~m is assumed. (c) The spatial profiles of $\Delta T$ near the heater at various $t$.}
		\label{Fig:1}
	\end{figure*}

In our study of transient heat transfer from a planar heater in He II, we consider a 1D computational domain with the heater located at $x=0$ as shown in Fig.~\ref{Fig:1}~(a). To generate a rectangular heat pulse of duration $t_h$ with a surface heat flux $q_h$, we adopt a time-dependent boundary condition on the heater surface: $v_n=q_h/\rho sT$ for the normal fluid and $v_s=-v_n\rho_n/\rho_s$ for the superfluid during $0<t\leq t_h$; and $v_n=v_s=0$ at $t>t_h$. An initial vortex-line density of $L_0=10^2$~cm$^{-2}$ is also assumed, which is comparable to typical densities of remnant vortices in He II containers~\cite{Awschalom-1984-PRL}. Indeed, the exact value of $L_0$ does not affect the simulation results for the range of $q_h$ considered in our work~\cite{SBaoTHTHeII}. We evolve Eqs.~(\ref{eq:mass})--(\ref{eq:L}) using the MacCormack's predictor-corrector scheme with a spatial step $\Delta x=10^{-5}$~m and a time step $\Delta t=10^{-8}$~s, and we also adopt a flux-corrected transport method to suppress numerical instabilities~\cite{fletcher-2003}.

To illustrate the key features of transient heat transfer in He II, we show in Fig.~\ref{Fig:1}~(b) the simulated spatial profiles of the temperature increment $\Delta T=T-T_b$ at various $t$ for a representative case with $q_h=30$ W/cm$^2$ and $t_h=0.295$ ms. This simulation was performed at a bath temperature $T_b=1.78$ K with a He II depth $H=1$~m. One can see that a second-sound pulse featured by a positive $\Delta T$ emerges in He II, which carries the heat energy and propagates away from the heater surface at the known second-sound speed (i.e., $c_2=19.6$ m/s at 1.78 K \cite{Donnelly-1998-JPCRD}). Near the heater surface where the vortex-line density $L$ grows rapidly, the interaction between the vortices and the second-sound pulse can convert the thermal energy carried by the pulse to locally deposited heat. This process results in a heated layer of He II (i.e., the thermal layer) adjacent to the heater surface~\cite{SBaoTHTHeII}. To make this thermal layer more visible, we plot the $\Delta T$ profile near the heater in Fig.~\ref{Fig:1}~(c). As the heat pulse ends, $\Delta T$ on the heater surface reaches the highest value. After that, the thermal layer spreads out diffusively~\cite{vansciver-2012} so that $\Delta T$ on the heater surface decreases. This example clearly shows that the largest temperature increment in He II occurs near the heater surface by the end of the heat pulse. Therefore, in our subsequent study on the onset of boiling in He II, we shall focus on the state variation of the He II adjacent to the heater.

\section{\label{sec:peak}Peak heat flux calculation}
\begin{figure}[t]
\centering
\includegraphics[width=0.95\linewidth]{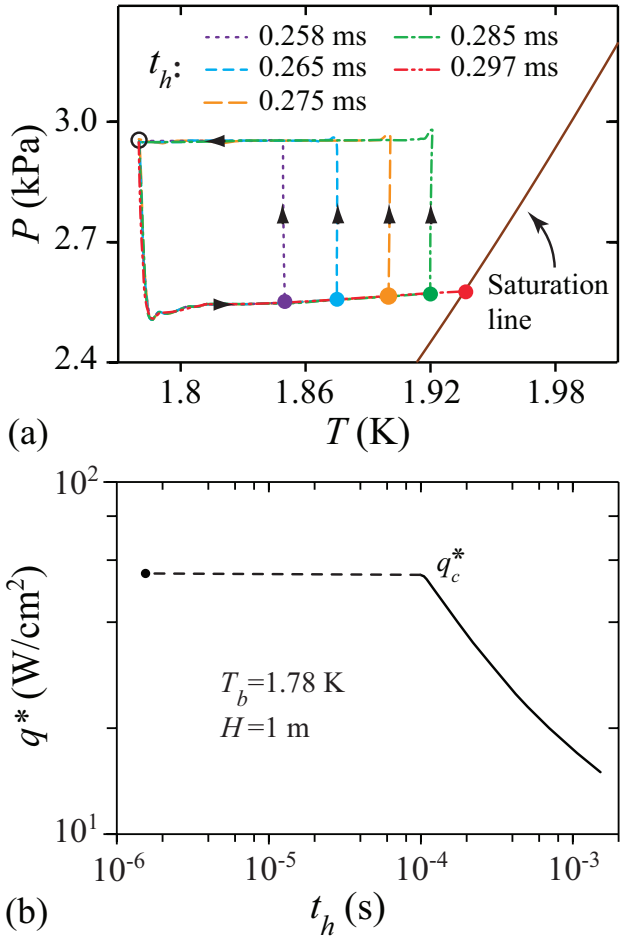}
\caption{(a) Evolution of the He II thermodynamics state at $x=\Delta x$ when heat pulses of $q_h=30$ W/cm$^2$ with different duration $t_h$ are applied. The filled circles indicate the He II states at the end of the heat pulses. (b) The simulated curve showing the dependance of the peak heat flux $q^*$ and the pulse duration $t_h$. The dashed line denotes that above a critical peak heat flux $q^*_c$, the onset time of boiling suddenly drops to the order of $\Delta x/c_2$.}
\label{Fig:2}
\end{figure}

For a given helium bath condition (i.e., $T_b$ and $H$), the peak heat flux $q^*$ depends on the heat-pulse duration $t_h$. To determine the correlation between $q^*$ and $t_h$, we adopt a method by scanning $t_h$ as illustrated in Fig.~\ref{Fig:2}~(a). This figure shows the evolution of the thermodynamic state $(P,T)$ of the He II at the first grid point $x=\Delta x$ for heat pulses with the same flux $q_h=30$~W/cm$^2$ but different duration $t_h$. All the curves start from the same initial state marked by the black open circle, i.e., $T|_{t=0}=T_b=1.78$~K and $P|_{t=0}=P_b=P_s(T_b)+\rho gH$ where $P_s(T_b)$ is the saturation pressure at $T_b$ and $H=1$~m. The He II states at the end of the heat pulses are marked by the filled circles of the respective colors. It is clear that when the heater turns on, there is a pressure drop followed by an increase of the He II temperature. As the heat pulse ends, the pressure spikes up in all the cases and the state curves evolve back to the starting point. Obviously, the He II state at the end of the heat pulse gets closer to the saturation line at larger $t_h$. We consider that boiling occurs when the state curve reaches the saturation line. For the example shown in Fig.~\ref{Fig:2}~(a), boiling occurs at $t_h=0.297$~ms.

By repeating the above analysis at various applied heat fluxes, we can determine the corresponding pulse durations beyond which boiling occurs. The results for $T_b=1.78$~K and $H=1$~m are collected in Fig.~\ref{Fig:2}~(b). We see that when $t_h$ is greater than about $10^{-4}$~s, $q^*$ increases with decreasing $t_h$, which agrees with the trend reported in literature~\cite{tsoi1985boiling,danilchenko1989limit,wang1995criterion,shimazaki1998measurement,kondaurova2017influence}. However, when the applied heat flux reaches a critical value $q^*_c\simeq55$~W/cm$^2$, we find surprisingly that the onset time of boiling suddenly jumps from about $10^{-4}$~s to an extremely small value. This value is found to be of the order $\Delta x/c_2$, i.e., the traveling time of the second-sound pulse to reach the first grid point. At heat fluxes higher than $q^*_c$, boiling always occurs on a similar time scale. This time scale $\Delta x/c_2$ suggests that the onset time for boiling would become arbitrarily small as one approaches the heater surface. However, in practice, the onset time will be limited by various factors such as the time it takes for vapor bubbles to grow on the heater surface, which is about a few microseconds for the bubbles to reach a radius of about 10~$\mu$m~\cite{Guo-2007-JLTP,Guo-2009-PRB}. The appearance of the critical peak heat flux $q^*_c$ and the associated sudden drop of the onset time of boiling to the order of $\Delta x/c_2$ are previously unreported phenomena, which indicates the existence of an unusual mechanism of boiling in He II.

\begin{figure}[t]
\centering
\includegraphics[width=0.95\linewidth]{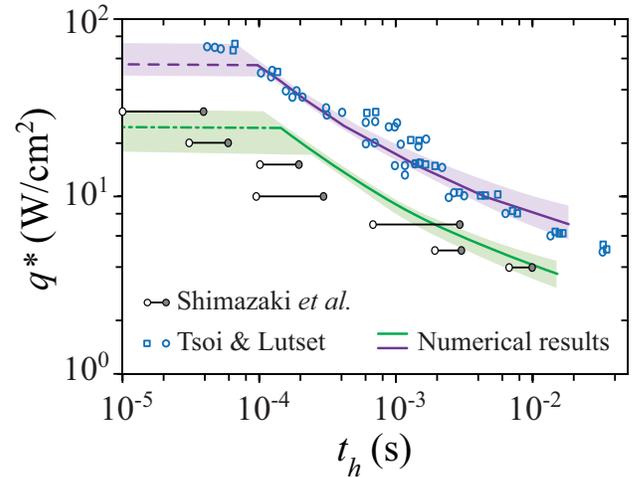}
\caption{Comparison of the simulated peak heat flux $q^*$ and some relevant experimental data. The blue circles and squares are data obtained by Tsoi and Lutset at 1.794~K~\cite{tsoi1985boiling}. The open and filled circles are data obtained by Shimazaki \emph{et al.} at 2~K~\cite{shimazaki1998measurement}, where the open circles indicate possible onset of the boiling while the filled circles denote firm observation of the boiling. The purple and the green curves are our simulation at $T_b=1.794$~K with $H=1$~m and $T_b=2$~K with $H=0.3$~m, respectively. The purple band indicates the span of the curve when $H$ is varied in the range of $0.8-1.5$~m, while the green band is for $H$ in the range of $0.2-0.4$~m.}
\label{Fig:3}
\end{figure}

Before moving to the next section to present our systematic study of $q^*_c$, we would like to compare our model simulations with some available experimental data. In Fig.~\ref{Fig:3}, we show the experimental data of $q^*$ versus $t_h$ obtained by Tsoi and Lutset~\cite{tsoi1985boiling} and by Shimazaki \emph{et al}~\cite{shimazaki1998measurement}. The work of Tsoi and Lutset adopted a thin-film nichrome heater (surface area: $3\times3$~cm$^2$) immersed in He II at 1.794~K, and the boiling was detected by monitoring the pressure change in He II using a piezosensor. The experiment of Shimazaki \emph{et al}. was conducted at $T_b=2$~K and utilized a slightly smaller heater (area: $2.7\times2.7$~cm$^2$), where the boiling was detected by measuring thermal shockwaves using a superconducting temperature sensor. The exact hydrostatic head pressures in these experiments were not reported. Nonetheless, we can perform simulations at the corresponding $T_b$ with a range of $H$ compatible with the expected He II depths estimated based on their setup schematics. The simulated $q^*$-$t_h$ curves are shown in Fig.~\ref{Fig:3}, which agree well with these experimental data. This agreement validates our model. Note that due to the limited sensor response times, the sudden drop of the onset time of boiling at heat fluxes above $q^*_c$ was not resolvable in these experiments.

\section{\label{sec:variation}Critical peak heat flux variation}
\begin{figure}[t]
	\centering
	\includegraphics[width=0.95\linewidth]{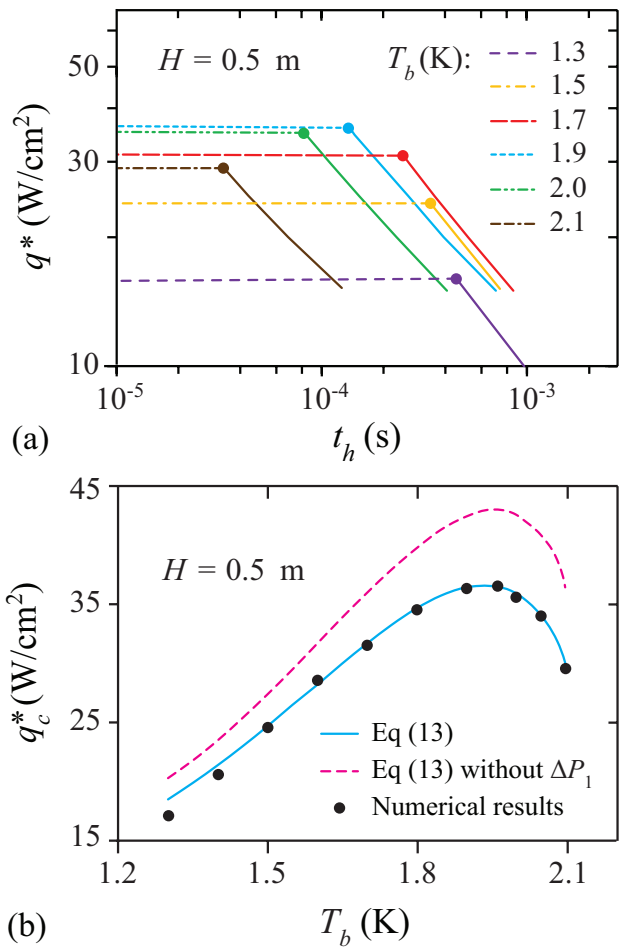}
	\caption{(a) Calculated $q^*$-$t_h$ curves at various $T_b$ with a fixed He II depth $H=0.5$~m. The critical peak heat flux $q^*_c$ for each curve is marked by the filled circle of the respective color. (b) The obtained critical peak heat flux $q^*_c$ as a function of $T_b$. The black dots are simulation data. The solid and the dashed curves are calculated using Eq.~(\ref{Eq:P-balance}) with and without the $\Delta P_1$ term, respectively.}
	\label{Fig:4}
\end{figure}

In order to understand the physical mechanism underlying the critical peak heat flux $q^*_c$, we need to first conduct a systematic study on its dependance on the helium bath condition. For this purpose, we have repeated the afore-mentioned analysis at various $T_b$ and $H$. In Fig.~\ref{Fig:4}~(a), we show the calculated $q^*$-$t_h$ curves for $T_b$ in the range of $1.3-2.1$~K with a fixed He II depth of $H=0.5$ m. The critical peak heat flux $q^*_c$ at each $T_b$ is identified and marked by the filled circle. From this study, the dependance of $q^*_c$ on $T_b$ is obtained, which is shown in Fig.~\ref{Fig:4}~(b). $q^*_c$ first increases with increasing $T_b$ before reaching a maximum at $T_b\simeq1.95$~K. Then, $q^*_c$ decreases as $T_b$ further increases. The maximum value of $q^*_c$ appears to be achieved at the bath temperature where the two fluids have about the same densities.

\begin{figure}[t]
	\centering
	\includegraphics[width=0.95\linewidth]{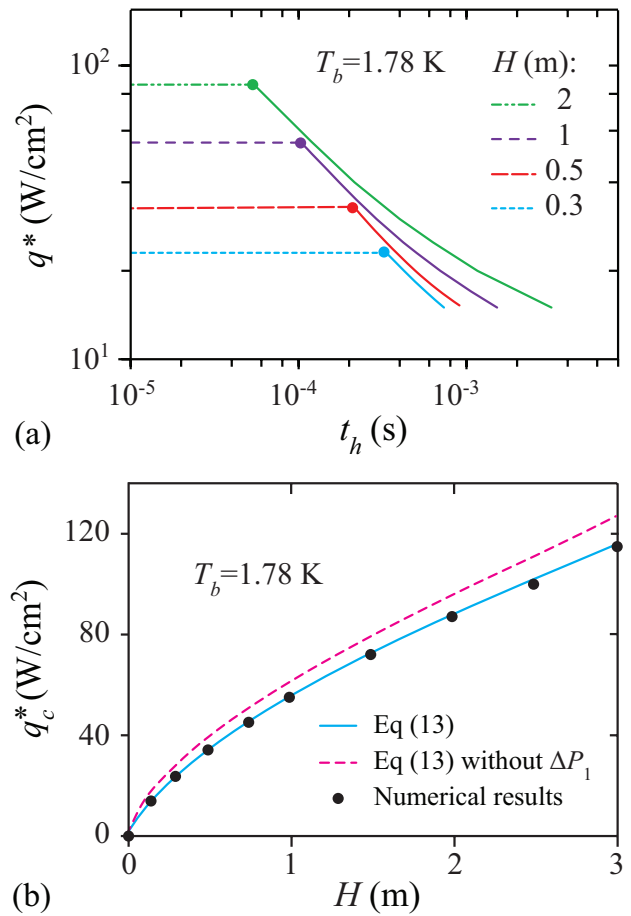}
	\caption{(a) Calculated $q^*$-$t_h$ curves at various $H$ with a fixed bath temperature $T_b=1.78$~K. The critical peak heat flux $q^*_c$ for each curve is marked by the filled circle of the respective color. (b) The obtained critical peak heat flux $q^*_c$ as a function of $H$. The black dots are simulation data. The solid and the dashed curves are calculated using Eq.~(\ref{Eq:P-balance}) with and without the $\Delta P_1$ term, respectively.}
	\label{Fig:5}
\end{figure}

When the He II depth $H$ is changed, the dependance of $q^*_c$ on $T_b$ remains similar to that presented in Fig.~\ref{Fig:4}~(b) but its exact value changes. To illustrate how $q^*_c$ varies with $H$ quantitatively, we fix the bath temperature at $T_b=1.78$~K and calculate the $q^*$-$t_h$ curves at various He II depth $H$. Representative results for $H$ in the range of $0.3-2$~m are shown in Fig.~\ref{Fig:5}~(a), where $q^*_c$ can be determined (marked by the filled circles). The obtained $q^*_c$ is then plotted as a function of $H$ in Fig.~\ref{Fig:5}~(b). It is clear that $q^*_c$ increases monotonically with increasing $H$.


\section{\label{sec:explanation}Explanation of critical peak heat flux}
\begin{figure}[t]
\centering
\includegraphics[width=0.95\linewidth]{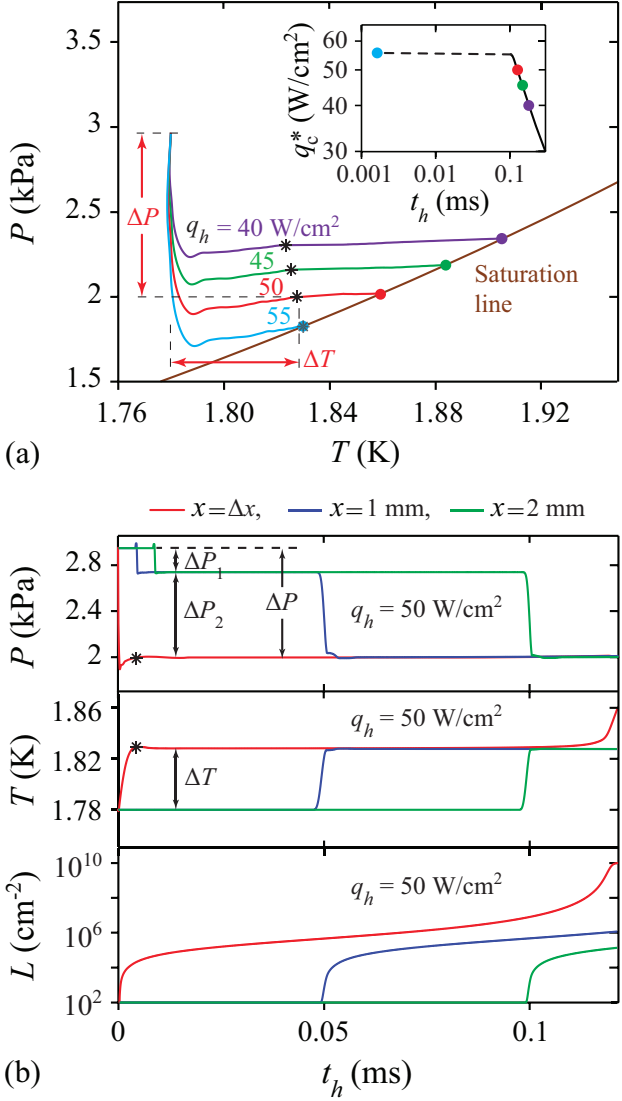}
\caption{(a) Evolution of the He II state at $x=\Delta x$ when heat pulses of different $q_h$ are applied. The asterisk in each curve denotes the end state of the fast process, and the filled circle marks where the state curve reaches the saturation line. The inset shows the obtained correlation between $q^*_c$ and $t_h$. (b) Time evolution of the He II pressure $P$, temperature $T$, and the vortex-line density $L$ at $x=\Delta x$, 1~mm, and 2~mm for the case with $q_h=50$~W/cm$^2$. All the simulations are conducted at $T_b=1.78$~K and $H=1$~m.}
\label{Fig:6}
\end{figure}

The studies presented in the previous sections show that the critical peak heat flux $q^*_c$ depends on both $T_b$ and $H$ and is likely associated with the propagation of the second-sound pulse since the corresponding boiling time is on the order of $\Delta x/c_2$. To better understand the physical processes that controls $q^*_c$, we show in Fig.~\ref{Fig:6}~(a) the evolution of the He II state adjacent to the heater (i.e., $x=\Delta x$) when the applied heat flux gradually increases. All the cases start from the same initial state as in Fig.~\ref{Fig:2}~(a), i.e., $T=T_b=1.78$~K and $P=P_s(T_b)+\rho gH$ with $H=1$~m. In what follows, we present a few important features observed in this study.

First, when the heater turns on, there is a fast process during which the pressure drops by $\Delta P$ and the temperature increases by $\Delta T$. This process occurs on a time scale of $\Delta x/c_2$. The end state of this fast process is marked by the asterisk for each curve in Fig.~\ref{Fig:6}~(a). It is clear that the magnitudes of both $\Delta P$ and $\Delta T$ increase with increasing the heat flux $q$. To provide a more direct view of the fast process, we show the time evolution of the He II pressure, temperature, and the vortex-line density at $x=\Delta x$, 1~mm, and 2~mm for a representative case with $q_h=50$~W/cm$^2$ in Fig.~\ref{Fig:6}~(b). For the $x=\Delta x$ curve, following the fast process the He II temperature and pressure remain nearly constant for over 0.1~s. During this period, the vortex-line density $L$ gradually grows following an initial rapid increase. This initial increase of $L$ is controlled by the generation term (i.e., the last term) in Eq.~(\ref{eq:L}), which is also the reason why the exact value of the initial line density $L_0$ does not affect the simulation result. When $L$ builds up to a sufficient level (i.e., of the order $10^8$~cm$^{-2}$), local heating due to the mutual friction makes the He II temperature rise again. This heating process drives the He II state at $x=\Delta x$ further towards the saturation line. Interestingly, at the critical peak heat flux $q^*_c\simeq55$~W/cm$^2$, the state curve of the He II at $x=\Delta x$ reaches the saturation line during the fast process without involving any subsequent slower heating process. Therefore, the onset time for boiling suddenly drops to the order of $\Delta x/c_2$ (see the inset in Fig.~\ref{Fig:6}~(a)). At heat fluxes higher than $q^*_c$, the boiling is largely controlled by the sudden drop in pressure across the saturation line, a phenomenon that is known as cavitation~\cite{Brennen-2013-book}.

\begin{figure}[t]
\centering
\includegraphics[width=0.85\linewidth]{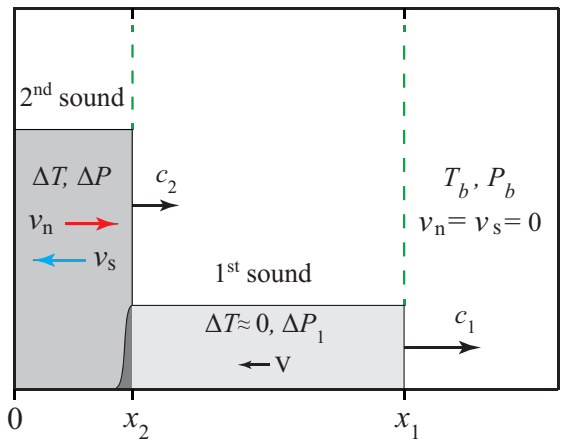}
\caption{A schematic diagram showing the temperature and pressure changes as well as the motion of the two fluids in the first-sound and the second-sound pulse zones. }
\label{Fig:7}
\end{figure}

Based on the physical picture presented above, we can indeed develop an analytical model to evaluate $q^*_c$. When the heater turns on, a second-sound pulse emerges from the heater surface. At short times when the vortex-line density near the heater is relatively low, the temperature increment $\Delta T$ within the second-sound pulse zone is related to the applied heat flux $q_h$ as~\cite{landau-1987}:
\begin{equation}
\begin{split}
q_h&= c_2T(\rho s|_{2nd}-\rho s|_{bath})\\
   &\simeq c_2\rho C_p\Delta T+\left.\frac{1}{2}v_{ns}^2c_2T\left(\frac{\partial\rho_n}{\partial T}\right)\right\vert_{T_b+\Delta T,P_b},
\label{Eq:q-T}
\end{split}
\end{equation}
where the subscripts ``2nd'' and ``bath'' denote the parameters evaluated in the second-sound pulse zone and in the He II bath, respectively. $C_p=T(\partial s/\partial T)_P$ is the specific heat of He II, and the second term in the above equation comes from the correction term in Eq.~(\ref{eq:s}). Within the second-sound pulse zone, a counterflow of the two fluids establishes as shown schematically in Fig.~\ref{Fig:7}. The velocities of the two fluids are given by $v_n$=$q_h/(\rho sT)|_{T_b+\Delta T}$ and $v_s$=$-v_n\rho_n/\rho_s$. Accordingly to Eq.~(\ref{eq:v}), the finite $v_n$ and $v_s$ lead to a pressure change $\Delta P_2$ in the second-sound pulse zone as given by:
\begin{equation}
\Delta P_2=-(\rho_sv_s^2+\rho_nv_n^2)\simeq-\left.\frac{q_h^2\rho_{n}}{s^2T^2\rho\rho_{s}}\right\vert_{T_b+\Delta T,P_b}\label{Eq:P2}.			
\end{equation}
This pressure drop is essentially a manifestation of the Bernoulli effect due to the motion of the two fluids in the second-sound zone.

Besides the effects due to the second-sound pulse, there is another subtle effect. Note that He II has a negative thermal expansion coefficient at temperatures above about 1.1~K~\cite{Donnelly-1998-JPCRD}. Therefore, the He II density $\rho$ in the second-sound zone must increase due to the temperature rise $\Delta T$, which requires a mass flow towards this region. To supply the mass, a first-sound pulse is generated where the two fluids move in phase at a velocity $\mathbf{v}$ towards the second-sound wavefront, as shown in Fig.~\ref{Fig:7}. The mass flux $\rho\mathbf{v}$ should balance the needed mass associated with the expansion of the second-sound pulse zone, i.e., $\rho|\mathbf{v}|=c_2(\rho|_{2nd}-\rho|_{1st})$. This finite $\rho\mathbf{v}$ leads to a pressure drop $\Delta P_1$ in the first-sound pulse zone. To the lowest order in $\mathbf{v}$, one can derive $\Delta P_1$ from Eq.~(\ref{eq:v}) as:
\begin{equation}
\begin{split}
&\Delta P_1\simeq-\rho c_1|\mathbf{v}|=-c_1c_2(\rho|_{2nd}-\rho|_{1st})\\
&\simeq-c_1c_2\left[(\rho|^{(s)}_{2nd}-\rho|^{(s)}_{1st})+\left.\frac{\rho^2v_{ns}^2}{2}\left(\frac{\partial\rho_n/\rho}{\partial P}\right)\right\vert_{2nd}\right]
\label{Eq:P1}
\end{split}
\end{equation}
where $c_1$ is the speed of the first sound in He II (i.e., about 231~m/s at 1.78~K). The pressure drop in the first-sound zone generates a pulling force exerting on the wavefront of the second-sound zone, which results in a total pressure drop in the second-sound zone as $\Delta P=\Delta P_1+\Delta P_2$. This effect can be clearly observed in Fig.~\ref{Fig:6}~(b) at $x=1$~mm (and $x=2$~mm), since the first-sound pulse and the second-sound pulse arrive at different times.

At the critical peak heat flux $q^*_c$, the temperature increment $\Delta T$ and the total pressure drop $\Delta P$ associated with the second-sound pulse would drive the He II at $x=\Delta x$ from the initial state ($T_b$,$P_b$) to the saturation line upon its arrival. Therefore, the following equation must hold:
\begin{equation}
P_s(T_b)+\rho gH+\Delta P_1+\Delta P_2=P_s(T_b+\Delta T)\label{Eq:P-balance},
\end{equation}
where $q^*_c$ enters the equation through $\Delta T$, $\Delta P_1$, and $\Delta P_2$ via Eqs.~(\ref{Eq:q-T})-(\ref{Eq:P1}). Using this model, we have calculated $q^*_c$ as a function $T_b$ at $H=0.5$~m and as a function of $H$ at $T_b=1.78$~K. The results are included in Fig.~\ref{Fig:4}~(b) and Fig.~\ref{Fig:5}~(b) as the solid curves. Excellent agreement between the model's calculation and the simulated $q^*_c$ values is observed, which proves that our understanding of the mechanism underlying the critical peak heat flux is correct. To see how large the first-sound effect is, we have also repeated the calculation using Eq.~(\ref{Eq:P-balance}) but without the $\Delta P_1$ term. The results are shown as the dashed curves in Fig.~\ref{Fig:4}~(b) and Fig.~\ref{Fig:5}~(b). Obviously, the first-sound effect is non-negligible at these large heat fluxes. 

\section{\label{sec:sum}Summary}
We have conducted a numerical study on 1D transient heat transfer in He II from a planar heater. The peak heat flux $q^*$ for the onset of boiling in He II is determined as a function of the heat-pulse duration $t_h$. A major finding in our study is the observation of a critical peak heat flux $q^*_c$ above which boiling occurs almost instantaneously. Our analysis shows that the boiling at heat fluxes lower than $q^*_c$ is caused by a heating process, which is associated with the relatively slow buildup of the quantized vortices and the thermal layer in front of the heater. When the applied heat flux is higher than $q^*_c$, the boiling is essentially a cavitation on the heater surface due to the combined effects of the first-sound and the second-sound waves in He II. A theoretical model for evaluating $q^*_c$ is developed, which nicely reproduces the simulated $q^*_c$ values at various He II bath temperatures and hydrostatic head pressures. Inspired by this work, a question we plan to address next is how the boiling physics may change in non-homogeneous heat transfer in He II. As shown in our early work~\cite{SBaoTHTHeII}, heat transfer of He II in non-homogeneous geometries (such as from cylindrical or spherical heaters) can exhibit new features. Understanding the behavior of the peak heat flux in these geometries could benefit research work such as quench-spot detection on He II cooled superconducting accelerator cavities~\cite{Bao-2019-PRApplied,Bao-2020-IJHMT} and the heat and mass transfer processes due to a vacuum failure in He II cooled accelerator beamline tubes~\cite{Garceau-2019-Cryo,Garceau-2019-IJHMT,Bao-2020-IJHMT-2,Garceau-2021-IJHMT}.

\section*{Acknowledgments}
The authors would like to acknowledge the support from the U.S. Department of Energy under Grant No. DE-SC0020113. The work was conducted at the National High Magnetic Field Laboratory at Florida State University, which is supported through the National Science Foundation Cooperative Agreement No. DMR-1644779 and the state of Florida.

\bibliography{Ref}

\end{document}